\documentclass[12pt]{iopart}
\usepackage{graphicx}
\usepackage{dcolumn}
\usepackage{bm}

\def\la{\mathrel{\mathpalette\fun <}}
\def\ga{\mathrel{\mathpalette\fun >}}
\def\fun#1#2{\lower3.6pt\vbox{\baselineskip0pt\lineskip.9pt
  \ialign{$\mathsurround=0pt#1\hfil##\hfil$\crcr#2\crcr\sim\crcr}}}

\begin{document}

\title[Magnetized Sources of Ultra-high Energy Nuclei and Origin of the Ankle]{Magnetized Sources of Ultra-high Energy Nuclei and Extragalactic Origin of the Ankle}

\author{G\"unter Sigl\dag, Eric Armengaud\dag}

\address{\dag
APC~\footnote[2]{UMR 7164 (CNRS, Universit\'e Paris 7, CEA, Observatoire de Paris)} (AstroParticules et Cosmologie),
11, place Marcelin Berthelot, F-75005 Paris, France\\
GReCO, Institut d'Astrophysique de Paris, C.N.R.S.,
98 bis boulevard Arago, F-75014 Paris, France}

\begin{abstract}
It has recently been suggested that ultra-high energy cosmic rays could
have an extragalactic origin down to the "second knee" at $\simeq4\times10^{17}\,$eV. In this case the "ankle" or "dip" at
$\simeq5\times10^{18}\,$eV would be due to pair production of extragalactic
protons on the cosmic microwave background which requires an injection
spectrum of about $E^{-2.6}$. It has been pointed out that for injection
of a mixed composition of nuclei a harder injection spectrum $\sim E^{-2.2}$
is required to fit the spectra at the highest energies and a galactic
component is required in this case to fit the spectrum below the ankle,
unless the proton fraction is larger than 85\%. Here we perform numerical
simulations and find that for sufficiently magnetized sources, observed spectra above $10^{19}\,$eV approach again the case of pure proton injection due to increased path-lengths and more efficient photo-disintegration of nuclei around the sources. This decreases secondary fluxes at a given energy and thus requires injection spectra $\sim E^{-2.6}$, as steep as for pure proton injection. In addition, the ankle may again be sufficiently dominated by protons to be interpreted as a pair production dip.
\end{abstract}

\pacs{98.70.Sa, 13.85.Tp, 98.65.Dx, 98.54.Cm}

{\bf Keywords}: ultra high energy cosmic rays, magnetic fields

\maketitle

\section{Introduction}
A major unresolved aspect of ultra-high energy cosmic ray (UHECR) physics~\cite{Cronin:2004ye} is their composition and above which energy the flux is dominated by extragalactic sources.
Above $\simeq10^{17}\,$eV the chemical composition is basically unknown~\cite{Watson:2004rg}. Around $10^{18}\,$eV the situation is particularly inconclusive as HiRes data~\cite{Abbasi:2004nz} suggest a light (proton dominated) composition, whereas other experiments indicate a heavy composition~\cite{Hoerandel:2004gv}.

As a consequence, there are currently two different scenarios: The "standard" one,
where a transition from a steeper, galactic heavy component to a flatter, extragalactic component dominated by protons takes place at the ankle at $\simeq5\times10^{18}\,$eV, see, e.g., Ref.~\cite{Wibig:2004ye},
and a more recent one suggesting that this transition actually takes place at lower energies, namely around the "second knee" at $\simeq4\times10^{17}\,$eV.

This second scenario in which extragalactic protons dominate down to the second knee has the following consequences: First, since the observed spectrum above the second knee is quite steep, $\propto E^{-3.3}$, the extragalactic proton flux has to cut off below $\simeq4\times10^{17}\,$eV. This can be explained as a magnetic horizon effect: Protons from cosmological distances cannot reach the observer any more within a Hubble time due to diffusion in large scale magnetic fields~\cite{Lemoine:2004uw,Aloisio:2004fz}.
Second, the ankle would have to be interpreted as due to pair production of the extragalactic protons~\cite{Berezinsky:2005cq}. In particular, it has been pointed out recently~\cite{Allard:2005ha}, that this model cannot afford injection of a significant heavy component above the ankle whose secondary photo-disintegration products in the form of intermediate mass nuclei would produce a bump around the ankle. Injection of a mixed composition would also require a harder injection spectrum
$\propto E^{-\alpha}$ with $\alpha\simeq2.2$, as opposed to the extragalactic ankle scenario with pure protons~\cite{Berezinsky:2005cq} which requires an injection spectrum $\propto E^{-2.6}$.

In the present study we point out that these conclusions can change significantly in the presence of large scale extragalactic magnetic fields (EGMF) which may immerse the UHECR sources. In fact, UHECR propagation simulations are usually performed for mixed compositions of nuclei without EGMF or
in the presence of EGMF, but restricted to pure protons. This is partly due to the increased difficulty posed by nuclei propagating in EGMF: Especially at lower energies, deflection can be considerable and require considerable CPU time. As a result, this most general case starts to being studied only since very recently~\cite{Armengaud:2004yt,Sigl:2004ff}.

We will apply our recent work to a mixed composition similar to the one inferred for Galactic sources, using the EGMF obtained from the large scale structure simulations performed in Ref.~\cite{Miniati:2002hs}. These EGMF are highly structured in that they reach a few microGauss in the most prominent structures such as galaxy clusters, but is $\la10^{-11}\,$G in the voids. The observer is chosen in a low magnetic field region with some resemblance to Earths actual environment. We note that other simulations suggest more concentrated EGMF~\cite{dolag} and current observations do not allow to distinguish between such different EGMF scenarios.

In section 2 we present our simulations, in section 3 we discuss implications for the interpretation of the ankle, and we conclude in section 4. We use natural units, $\hbar=c=k=1$, throughout the paper.

\section{Simulations above $10^{19}\,$eV}

We restrict simulations to energies $\geq10^{19}\,$eV because of CPU time limitations due to strong deflection at lower energies and due to lack of resolution in the simulated EGMF.
We use the same approach as described in all detail in Ref.~\cite{Armengaud:2004yt}.
Specifically, we use a source density of $2.4\times10^{-5}\,{\rm Mpc}^{-3}$, assuming the (astrophysical) sources~\cite{Torres:2004hk,Bhattacharjee:1998qc} follow the baryons. We inject a power law up to maximal rigidity $E/Z$ of $4\times10^{20}\,$eV, whose index $\alpha$ can be fitted to match the observed spectrum after propagation. To take into account cosmic variance due to fluctuations in source positions and properties we simulate typically about 60 realizations for the discrete source distribution. In addition, for each such realization, we dial 20 realizations for the source power which we let fluctuate by a factor 100 and for the injection spectral index which we let fluctuate by 0.1 around the average fitted $\alpha$, as in Ref.~\cite{Armengaud:2004yt}. For propagation, all relevant energy loss processes are implemented, including pair production, pion production, and photo-disintegration on the combined radio, cosmic microwave, and infrared background for all nuclei~\cite{Bertone:2002ks}. In contrast to Ref.~\cite{Armengaud:2004yt} we now consider a more realistic mixed composition at injection that we define as follows:

\begin{figure}[ht]
\includegraphics[width=0.6\textwidth,clip=true]{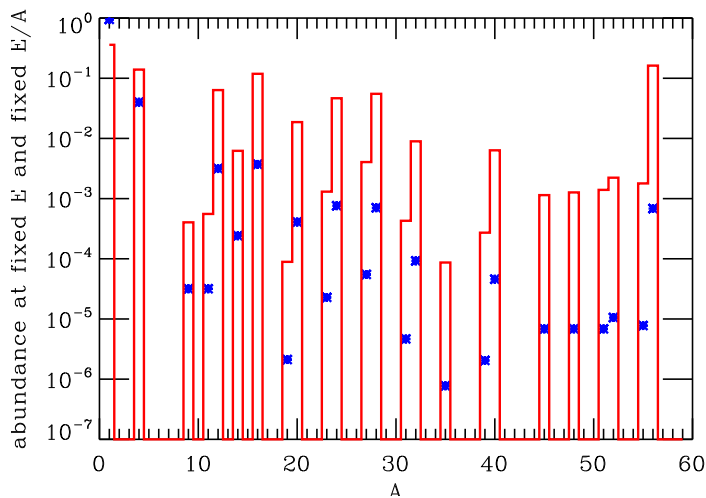}
\caption{Elemental abundances at a given energy per nucleon (blue asterisks) and resulting abundances at a given total energy for an injection spectrum $\propto E^{-2.6}$ (histogram).}
\label{fig1}
\end{figure}

Let the abundance of a nucleus of atomic mass $A$ at a given energy per nucleon
be $x_A$. If the spectra of all nuclei have a common slope $\propto E^{-\alpha}$,
the injected differential spectrum $(dn_A/dE)(E)$ for species $A$ at a given total energy $E$ is then given by
\begin{equation}
  \frac{dn_A}{dE}(E)\simeq N\,x_A\,A^{\alpha-1}\,E^{-\alpha}\,,\label{spec_A}
\end{equation}
where $N$ is a normalization constant. The modified abundance factors result from conversion from $E/A$ to $E$. In the present work we will assume
$x_1=95\%$ for the proton fraction, $x_4=4\%$ for the helium fraction,
see Ref.~\cite{Swordy:1993}. For the remaining 1\% nuclei with $A>4$ we follow
Ref.~\cite{Allard:2005ha} in adopting the distribution from Ref.~\cite{DuVernois:1996}.
The abundance of a stable element $^AZ$ in the Galactic cosmic ray spectrum will
be proportional to the product of the injection flux Eq.~(\ref{spec_A}) and the Galactic confinement time $\tau(E/Z)$ which below the knee approximately scales as $(E/Z)^{-0.6}$. Since Galactic sources are inferred to have an injection spectral index $\alpha\simeq2.1$, Galactic cosmic ray abundances below the knee will be roughly $\propto x_A\,A^{1.1}\,Z^{0.6}$. With the choices for $x_A$ above, this results in abundances at fixed total energy that are consistent with observations in the knee region~\cite{Hoerandel:2004gv,Swordy:1993} within a factor 2.

Below we will find that significantly magnetized extragalactic sources should
inject a spectrum roughly $\propto E^{-2.6}$ to fit the data above $10^{19}\,$eV.
The abundances $x_A$ at a given energy per nucleon and the ones at a given total energy resulting from Eq.~(\ref{spec_A}) in this case are shown in Fig.~\ref{fig1}.

In the absence of other interactions, if nuclei of mass $A$ at energy $AE/A^\prime$ are disintegrated into nuclei of mass $A^\prime$ and energy $E$, then we have for the flux $(dn_{A^\prime}/dE)(E)$ of these secondary nuclei
\begin{equation}
  \frac{dn_{A^\prime}}{dE}(E)\simeq   \left(\frac{A}{A^\prime}\right)^2\frac{dn_A}{dE}\left(\frac{A}{A^\prime}E\right)
  \simeq N\,x_A\,A\,\,{A^\prime}^{\alpha-2}E^{-\alpha}
  \,.\label{spallation}
\end{equation}  
where in the second step we have used Eq.~(\ref{spec_A}) for the differential
flux of the parent nuclei. This formula holds trivially also for the primary proton flux for $A=A^\prime=1$.

\begin{figure}[ht]
\includegraphics[width=0.6\textwidth,clip=true]{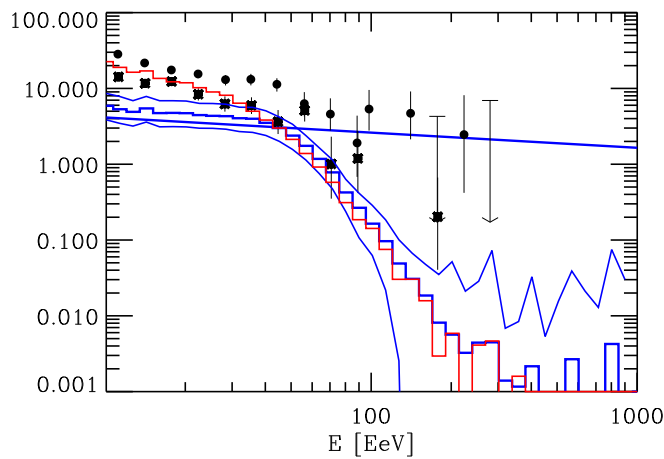}
\caption{Realization averaged observable all-particle spectrum for injection with spectrum $\propto E^{-2.2}$ (solid horizontal line) and mixed composition as discussed in the text, without (red histogram) and with (blue histogram) EGMF. Cosmic variance (lines around blue histogram) is only shown for the case with EGMF. Also shown are some data from the AGASA~\cite{agasa} and HiRes~\cite{hires} experiments.}
\label{fig2}
\end{figure}

\begin{figure}[ht]
\includegraphics[width=0.6\textwidth,clip=true]{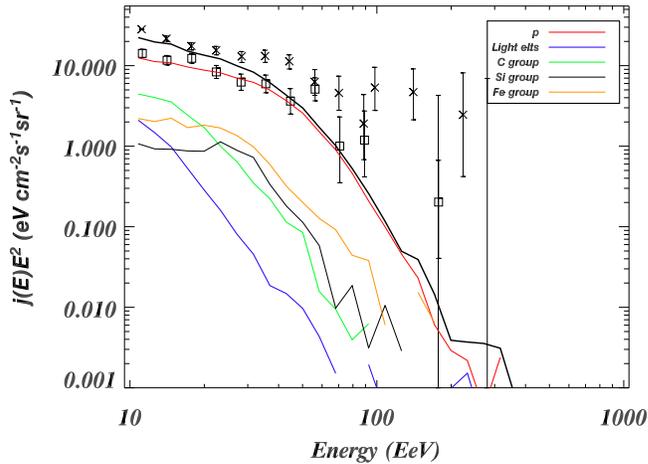}
\caption{Average over source realizations of the all-particle spectrum (top
line) and of spectra for different mass groups in the case of injection of a mixed composition with spectrum $\propto E^{-2.2}$ in the absence of EGMF.
For particles heavier than nucleons, we have made the
following groups for convenience: ``Light elements'' (blue): $2 \le
A \le 11$; ``C group'' (green): $12 \le A \le 24$; ``Si
group'' (black): $25 \le A \le 40$; ``Fe group'' (yellow): $41 \le
A \le 56$.}
\label{fig3}
\end{figure}

Photo-disintegration leads to a steepening of the spectrum compared to
the injection spectrum. Eq.~(\ref{spallation}) shows that at a given energy incomplete spallation into nuclei $A^\prime >1$ tends to lead to secondary fluxes that are higher than for complete spallation into protons. The enhancement factor is given by ${A^\prime}^{\alpha-2}$ multiplied by the sum of $x_A\,A$ over the relevant parent nuclei for $A^\prime$ divided by the same factor for protons. As a consequence, at energies around $10^{19}\,$eV where photo-disintegration starts to become negligible, the steepening is only significant for incomplete
spallation. Fig.~\ref{fig2} shows that this is the case for negligible EGMF, but not in presence of considerable source magnetization. A similar but more moderate difference between magnetized and unmagnetized sources was already seen in Ref.~\cite{Armengaud:2004yt} for pure iron injection. As a consequence, in the absence of EGMF, a relatively hard injection spectrum $\propto E^{-2.2}$ is required to fit the spectrum above $10^{19}\,$eV.
If we were to use a softer injection spectrum $\propto E^{-2.6}$, we would not fit the observed slope above $10^{19}\,$eV and we would predict a proton fraction around $10^{19}\,$eV of $\simeq27\%$, too small to explain the ankle. These findings are consistent with Ref.~\cite{Allard:2005ha} who used a homogeneous source distribution. 
We have verified that in the absence of EGMF and for the source density we use, there is no significant difference in the results between structured and homogeneous source distributions.

\begin{figure}[ht]
\includegraphics[width=0.6\textwidth,clip=true]{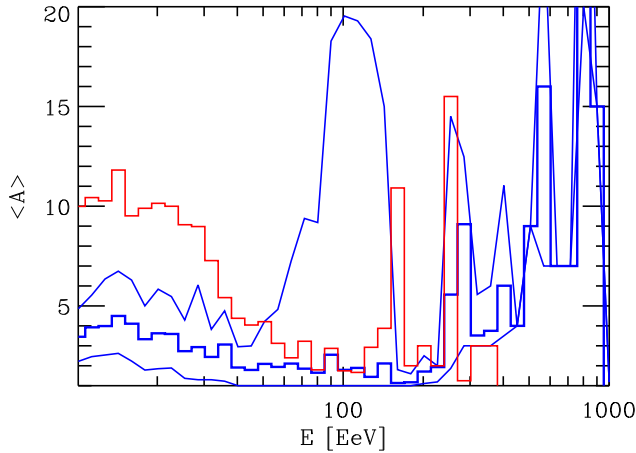}
\caption{Realization averaged observable composition for injection with spectrum $\propto E^{-2.2}$ and mixed composition as discussed in the text, without (red histogram) and with (blue histogram) EGMF. Cosmic variance (lines around blue histogram) is only shown for the case with EGMF.}
\label{fig4}
\end{figure}

Fig.~\ref{fig3} resolves the observable spectrum into individual mass group elements for an injection spectrum $\propto E^{-2.2}$ in the absence of EGMF. The proton fraction at $10^{19}\,$eV is $\simeq58\%$.

Fig.\ref{fig4} shows that, for a fixed injection spectrum, increased spallation in EGMF can indeed lead to a lighter observable composition around $10^{19}\,$eV.

\begin{figure}[ht]
\includegraphics[width=0.6\textwidth,clip=true]{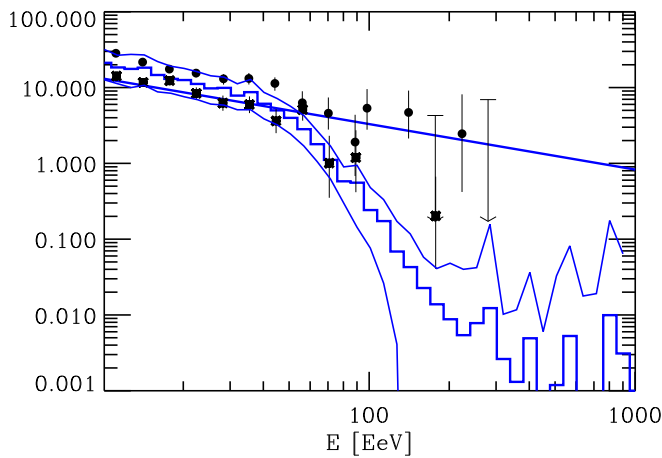}
\caption{Realization averaged observable all-particle spectrum for injection with spectrum $\propto E^{-2.6}$ (solid line) and mixed composition as discussed in the text, in presence of EGMF. Line key as in Fig.~\ref{fig2}.}
\label{fig5}
\end{figure}

\begin{figure}[ht]
\includegraphics[width=0.6\textwidth,clip=true]{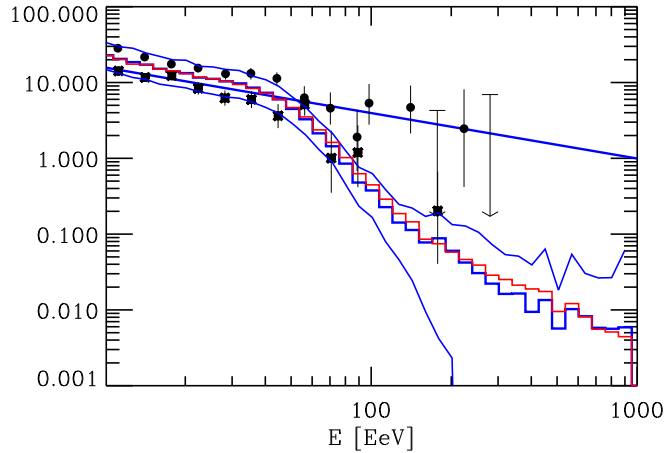}
\caption{Same as Fig.~\ref{fig5}, but for pure proton injection,
with (blue histogram) and without (red histogram) EGMF.}
\label{fig6}
\end{figure}

Enhanced, more complete spallation in the EGMF can lead to more protons which
follow more closely the original injection spectrum because their secondary flux is smaller than for incomplete spallation, as explained above. This can be seen in Fig.~\ref{fig2}. A similar tendency was already seen in Ref.~\cite{Armengaud:2004yt} for pure iron injection. As a result, a steeper injection spectrum has to be chosen to fit the spectrum observed above $10^{19}\,$eV, as demonstrated in Fig.~\ref{fig5}. We note that around $10^{19}\,$eV the case of a mixed composition fits better than pure iron injection studied in Ref.~\cite{Armengaud:2004yt}. This is because pure iron injection gives rise to a bump at a few $10^{19}\,$eV, somewhat flattened out by the EGMF.

\begin{figure}[ht]
\includegraphics[width=0.6\textwidth,clip=true]{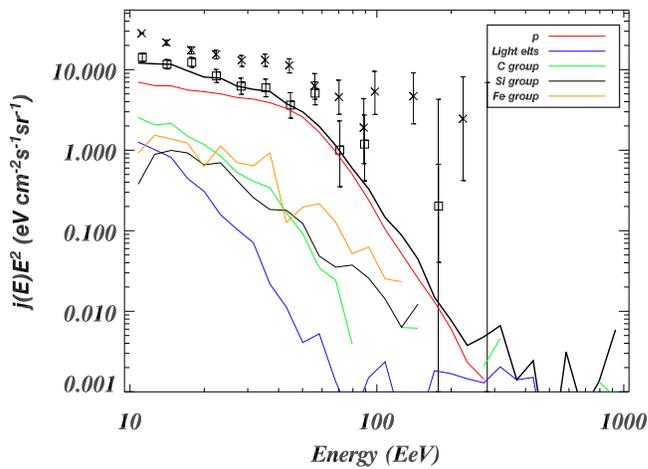}
\caption{Same as Fig.~\ref{fig3}, but for injection of the mixed composition with spectrum $\propto E^{-2.6}$ in the presence of EGMF.}
\label{fig7}
\end{figure}

The more complete spallation in case of magnetized sources consequently leads to a spectrum closer to the pure proton injection case, see Fig.~\ref{fig6}. Note that in the case of pure proton injection, the observable spectrum is hardly influenced by EGMF. The spectrum is in fact close to the "universal spectrum" which is independent of the propagation mode and applies if particle number is conserved and if sources are distributed roughly uniformly with separations small compared the characteristic propagation lengths, namely energy loss and diffusion lenghts~\cite{Aloisio:2004jd}. Note that, although highly structured on scales $\la10\,$Mpc, our source distribution is roughly uniform on larger scales, with typical source separations of $\simeq30\,$Mpc. In contrast, the universal spectrum is not applicable in case of mixed compositions since the number of cosmic rays is then no longer conserved. Indeed, the observable spectrum does depend considerably on EGMF in this case even taking into account considerable cosmic variance, as we have seen in Fig.~\ref{fig2}. A numerical approach is, therefore, indispensable in this case.

\begin{figure}[ht]
\includegraphics[width=0.6\textwidth,clip=true]{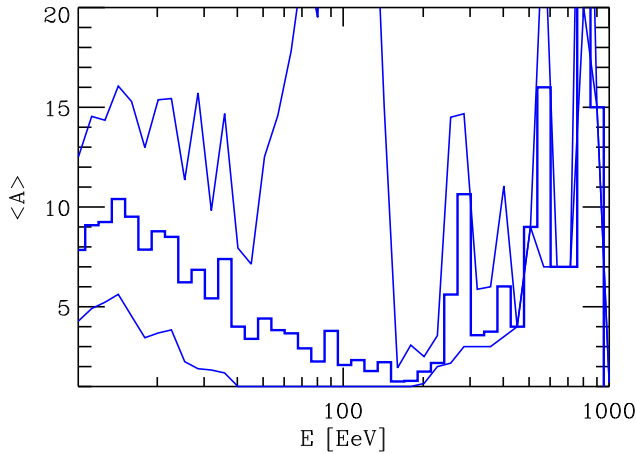}
\caption{Observable composition for injection with spectrum $\propto E^{-2.6}$
and mixed composition as discussed in the text, in presence of EGMF. Line key as in Fig.~\ref{fig4} for presence of EGMF with harder injection spectrum.}
\label{fig8}
\end{figure}

Fig.~\ref{fig7} resolves the observable spectrum into individual mass group elements for an injection spectrum $\propto E^{-2.6}$ in the presence of EGMF. The proton fraction at $10^{19}\,$eV is $\simeq63\%$.

In the presence of EGMF the composition around $10^{19}\,$eV is heavier for a steeper injection spectrum, see Fig.~\ref{fig8}, than the corresponding one shown
in Fig.~\ref{fig4} for a harder injection spectrum. This is mostly due to the $A^{\alpha-1}$ enhancement of heavy elements for a given total energy, see Eq.~(\ref{spec_A}).

\section{The Ankle}

In the previous section we saw that magnetized sources of a mixed composition require injection indices $\alpha\simeq2.6$ to fit the spectrum above $10^{19}\,$eV, very similar to pure proton injection, and contrary to unmagnetized sources with $\alpha\simeq2.2$. The latter spectrum is certainly too hard to explain the ankle and would underproduce the observed flux below a few times $10^{18}\,$eV where the injection spectrum is not modified by interactions anymore. In contrast, the magnetized scenario may still explain the ankle as a pair production feature of an extragalactic proton population dominating at that energy, as we are now going to argue.

\begin{figure}[ht]
\includegraphics[width=0.6\textwidth,clip=true]{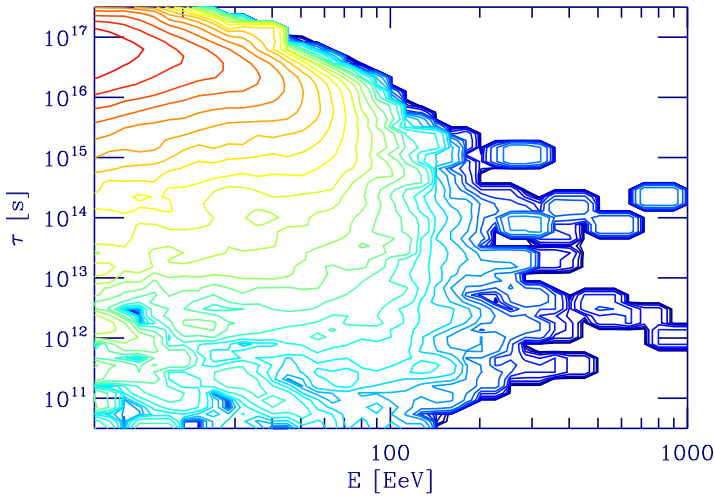}
\caption{Histogram of delay times versus observed energy for the simulation with injection of a mixed composition with spectrum $\propto E^{-2.6}$ in the presence of EGMF. The contour lines are logarithmic in steps of 0.25 dex.}
\label{fig9}
\end{figure}

The simulations show that confinement (delay) times at energies $E\simeq10^{19}\,$eV are $\sim5\times10^{16}\,$s, see Fig.~\ref{fig9}. This is indeed consistent with an analytical estimate of the confinement time of light nuclei of energy $E$ in large scale structures (galaxy clusters) of size $R\sim\,$Mpc with magnetic fields $B\sim\mu\,$G,
\begin{equation}
  \tau(E)\sim\frac{R^2}{6D(E)}\sim\frac{ZeBR^2}{2E}
  \sim4\times10^{16}\,Z\left(\frac{B}{\mu\,{\rm G}}\right)
  \left(\frac{R}{3\,{\rm Mpc}}\right)^2\left(\frac{10^{19}\,{\rm eV}}{E}\right)
  \,{\rm s}\,,\label{t_delay}
\end{equation}
assuming Bohm diffusion, $D(E)\simeq E/(3ZeB)$. Below $10^{19}\,$eV the simulations cannot be used because the Larmor radius of charged particles becomes smaller than the resolution of the magnetic field in the simulation. However, extrapolating Eq.~(\ref{t_delay}) to smaller energies provides us with conservatively long confinement times given that Bohm diffusion gives the smallest possible diffusion coefficient. Confinement times of light nuclei are, therefore, smaller than the age of the Universe down to $\simeq10^{18}\,$eV and the cosmic rays of energy $10^{18}\,{\rm eV}\la E\la 10^{19}\,$eV will leave the strongly magnetized regions around the sources with roughly their original injection spectrum. Outside these magnetized regions space is weakly magnetized and the conditions for the universal spectrum should be fulfilled in this energy range~\cite{Aloisio:2004fz,Aloisio:2004jd}. In particular, the ankle should be reproduced as an extragalactic feature in this case, provided that the spectrum is sufficiently dominated by protons at these energies.

In order to address the composition below $\simeq10^{19}\,$eV, let us now assume
complete spallation above $\simeq10^{19}\,$eV, and no spallation below. The protons produced by spallation will pile up around $10^{19}\,$eV.
Then in some range of energies around the ankle the proton flux will be
given by the sum of Eq.~(\ref{spallation}) over all $A$ for $A^\prime=1$.
On the other hand,
the flux $(dn_h/dE)(E)$ of all nuclei heavier than nucleons which are not
disintegrated at these energies is given by the sum of Eq.~(\ref{spec_A})
over all $A>1$. For the fraction of protons we thus obtain the rough estimate,
\begin{equation}
  \frac{(dn_p/dE)(E)}{(dn_h/dE)(E)}\sim
  \frac{\sum_{A\geq1}x_A\,A}{\sum_{A>1}x_A\,A^{\alpha-1}+\sum_{A\geq1}x_A\,A}
  \,.\label{p_frac}
\end{equation}
If spallation above $10^{19}\,$eV is incomplete, the proton fraction
around the ankle will be smaller, since the numerator in Eq.~(\ref{p_frac}) decreases
towards $x_1$ and the denominator increases due to the contribution of intermediate mass nuclei produced by spallation of nuclei above $10^{19}\,$eV. This tendency is confirmed by the full simulations for an injection spectrum $\propto E^{-2.2}$ which give a proton fraction at $10^{19}\,$eV of $\simeq84\%$ with EGMF and $\simeq58\%$ without EGMF, compared to $\simeq69\%$ from Eq.~(\ref{p_frac}).

For $\alpha=2.6$, obtained from fitting the spectrum above $10^{19}\,$eV in the presence of EGMF, see previous section, we obtain $\simeq44\%$ for the proton fraction estimated from the simple analytical argument Eq.~(\ref{p_frac}). The full simulations from Fig.~\ref{fig7} gave a proton fraction of $\simeq63\%$ at $10^{19}\,$eV, tending to grow with decreasing energy, and the fraction of the heavier mass groups is $\la20\%$, and that of iron $\la10\%$. Within the uncertainties this is probably sufficiently high to explain the ankle by pair production of dominantly protons. In fact, Ref.~\cite{Allard:2005ha} obtained that $\ga85\%$ proton fraction at a given observed energy is needed, and from Ref.~\cite{Wibig:2004ye} one may read off that the heavy (iron) fraction should be $\la10\%$. 

Finally, at energies below the energy per nucleon where photo-disintegration lengths become comparable to the age of the Universe, i.e. for $E\la 3\times10^{18}\,$eV, the observed composition should approach the one injected at a given total energy. For the magnetized case this composition was shown in Fig.~\ref{fig1} as the red histogram.

\section{Conclusions}
Injection of a mixed extragalactic ultra-high energy cosmic ray population with composition similar to the one inferred for Galactic sources requires a relatively soft injection spectrum $\propto E^{-2.6}$ if sources are significantly magnetized. This is in contrast to the case of negligible magnetic fields which requires harder injection spectra $\propto E^{-2.2}$ because of increased secondary fluxes of lower energy light nuclei produced in incomplete photo-disintegration of heavy nuclei. Whereas the latter case requires a Galactic component below the ankle, the magnetized case with mixed composition may still explain the ankle as pair production of an extragalactic component dominated by protons and light nuclei piling up around $10^{19}\,$eV. This light component dominates the flux around the ankle and is caused by increased spallation of the injected mixed composition due to considerable deflection around the sources. At energies $E\la 3\times10^{18}\,$eV the flux may still be extragalactic, but the observed composition should approach the one injected at a given total energy, and may thus have a significant heavy component.

If one believes that UHECR are produced by relativistic shock acceleration favoring spectral indices $\alpha\simeq2.2$~\cite{Achterberg:2001rx},
then injection of a mixed composition around considerably magnetized sources is disfavored. This argument may not, however, be very strong since there are indications that the spectral index for relativistic shock acceleration may not be universal~\cite{Baring:2004qc}.

Our results provide yet another example how structured large scale magnetic fields can modify spectra and scenarios not only for discrete ultra-high energy cosmic ray sources~\cite{Sigl:2004ff}, but also for the cosmological sources of the diffuse flux. In addition, the impact of magnetic fields on the spectrum depends considerably on the injected composition: For pure proton injection, the effect is insignificant and the spectrum is close to the universal spectrum for small source separations. In contrast, for a mixed composition close to the one inferred for Galactic sources considered in the present work, magnetic fields can modify the spectrum considerably, and even stronger than for pure iron injection.

We also stress that observation of a heavy composition $\langle A\rangle\ga 15$ above $\simeq10^{19}\,$eV would be almost impossible to explain by standard scenarios involving extragalactic sources at distances of tens of Mpc injecting compositions comparable to the ones inferred for Galactic sources.

\ack
This work partly builds on earlier collaborations with Gianfranco Bertone, Torsten En\ss lin, Claudia Isola, Martin Lemoine, and Francesco Miniati. We acknowledge discussions with Denis Allard.

\end{document}